\documentclass[lettersize,journal]{IEEEtran}
\usepackage{ifpdf}
\usepackage{booktabs}

\usepackage{cite}

\ifCLASSINFOpdf
\usepackage[pdftex]{graphicx}
\else
\usepackage[dvips]{graphicx}
\fi
\usepackage{amsmath}
\usepackage{algorithmic}

\usepackage{array}

\ifCLASSOPTIONcompsoc
\usepackage[caption=false,font=normalsize,labelfont=sf,textfont=sf]{subfig}
\else
\usepackage[caption=false,font=footnotesize]{subfig}
\fi
\usepackage{fixltx2e}

\usepackage{stfloats}
	\ifCLASSOPTIONcaptionsoff
	\usepackage[nomarkers]{endfloat}
	\let\MYoriglatexcaption\caption
	\renewcommand{\caption}[2][\relax]{\MYoriglatexcaption[#2]{#2}}
	\fi
	\usepackage{url}

	\usepackage{bm}
	\usepackage{enumitem}
	\usepackage{color}
	\usepackage{soul} 
	\usepackage{color, xcolor} 
	\renewcommand{\figurename}{Fig. }
	\usepackage{makecell}
	\usepackage{multirow}

\begin{document}
		%
		\title{ Transient Synchronization Stability Analysis and \\ Assessment of DFIG System \\ Under Severe Faults}
		%
		%
		\author{Hongsheng~Xu,
			Meng~Zhan,~\IEEEmembership{Senior Member,~IEEE}
			\thanks{
				This work was supported by the National Natural Science Foundation of China, grant number U22B6008. {\it{(Corresponding author: Meng Zhan.)}}
				
				Hongsheng Xu and Meng Zhan are with the State Key Laboratory of Advanced Electromagnetic Engineering and Technology, Hubei Electric Power Security and High Efficiency Key Laboratory, School of Electrical and Electronic Engineering, Huazhong University of Science and Technology, Wuhan 430074, China (e-mail: xhs@hust.edu.cn;~zhanmeng@hust.edu.cn).
					
			}
		}

	\maketitle
	\begin{abstract}
In the transient stability analysis of renewable energy grid-tied systems, although a large amount of works have devoted to the detailed electromagnetic transient simulation and the stability analyses of during-fault stage, the whole low-voltage ride through (LVRT) process and relevant transient stability mechanism remain to be uncovered. Taking the doubly fed induction generator system as the objective, this paper  divides the transient processes into four different stages, including the pre-fault, during-fault, early post-fault, and late post-fault ones, establishes the full mechanism models for each stage, and studies the switching dynamics in detail. It is found that the during-fault dynamics can be determined by the phase-lock loop second-order equation within the framework of the generalized swing equation (GSE). For the early post-fault stage, it can be treated as a series of quasi-steady states and its dominant driving system dynamics can still be described by the GSE. Based on the local dynamics of unstable equilibrium point, the system transient stability can be completely determined by whether the initial state of the early post-fault stage is within or out of its basin of attraction (BOA). Based on these observations, the BOA-based and equal area criterion (EAC)-based transient stability assessment methods are developed, which are supported by broad numerical simulations and hardware-in-the-loop experiments. This work provides a clear physical picture and perfectly solves the difficult stability analysis problem when severe faults and LVRT have to be considered in most of DFIG engineering situations.
	\end{abstract}


	%
	\IEEEpeerreviewmaketitle

	\section*{Nomenclature}
	\begin{itemize}[leftmargin = 63pt]
		\item[$\bm{U}_{\rm{\mathit{t}}}$, $\bm{I}_{\rm{\mathit{t}}}$, $\bm{U}_{\rm{\mathit{g}}}$] Terminal voltage and output current and grid voltage vectors.
		\item[$\bm{I}_{\rm{\mathit{s}}}$, $\bm{I}_{\rm{\mathit{t}}}$] Stator and rotor currents.		
		\item[${U}_{\rm{\mathit{t}}}$,  ${U}_{\rm{\mathit{g}}}$] Amplitude of the terminal voltage and grid voltage.
		\item[$i_{\rm\mathit{td}}$, $i_{\rm\mathit{tq}}$] $dq$ axis components of output current of DFIG.


		\item[$u_{\rm \mathit{td}}$, $u_{\rm\mathit{tq}}$] $dq$ axis components of terminal voltage.
		\item[$\theta _{\rm \mathit{pll}}$ ] Phase-locked loop (PLL) output angle in three-phase stationary $abc$ reference frame.
		\item[$\varphi _{\rm \mathit{pll}}$ ] PLL output angle in $xy$ common reference frame.
		\item[${\varphi_{\rm\mathit{cr}}}$] Critical clearing angle. 
		
		\item[$k_{\rm \mathit{pw}}$, $k_{\rm \mathit{iw}}$]  PI parameters of RSC.
		\item[$k_{\rm \mathit{pV}}$, $k_{\rm \mathit{iV}}$]  PI parameters of TVC.
		\item[$k_{\rm \mathit{ppll}}$, $k_{\rm \mathit{ipll}}$]  PI parameters of PLL.
		\item[$K_{\rm \mathit{e}}$, $K_{\rm \mathit{ramp}}$]  Reactive current ratio coefficient and ramp rate.

		\item[${P}_{\rm{\mathit{t}}}$,  ${Q}_{\rm{\mathit{t}}}$]  Output active power and reactive power of DFIG.

		\item[$\omega_0$] Rotation speed of $xy$ common reference frame.
		\item[$\omega _{\rm \mathit{pll}}$, $\omega _{\rm \mathit{g}}$, $\omega _{\rm \mathit{r}}$] Frequency of PLL and grid, and rotor speed, respectively.
		
		\item[$X_\mathit{m}$, $X_\mathit{s}$] Mutual reactance and stator reactance.
		
		\item[$a$, $b$, $c$, $d$] Correction coefficients.
		
		\item[${\bf{x}}_s$, ${\bf{x}}_u$] Stable and unstable equilibrium points.

		\item[${P_{\rm{\mathit{m}}}}$,${P_{\rm{\mathit{e}}}}$,${M_{\rm{\mathit{eq}}}}$,${D_{\rm{\mathit{eq}}}}$] Equivalent mechanical power, electromagnetic power, inertia, and damping, respectively.
		
		\item[$1,2,3,4$] Subscripts of stages 1, 2, 3, and 4, for pre-fault, during-fault,early post-fault, and late post-fault, respectively.
		
	\end{itemize}

	\section{Introduction}
	%
	%
	%
	%
	\IEEEPARstart{I}{n} recent years, doubly fed induction generator (DFIG) has become a mainstream renewable energy equipment in power systems \cite{Ref_3}. Compared to the synchronous generator (SG), the DFIG exhibits insufficient over-current capacity under severe faults, and many countries have developed grid codes for the operation of DFIG to avoid its off-grid \cite{Ref_6},\cite{Ref_GB}. Based on these grid codes, the DFIG should experience multiple switching processes during the low-voltage ride through (LVRT). The relevant transient synchronous stability (TSS) analysis and assessment has become a hot topic \cite{Ref_4, Ref_5, Ref_ZYY}.

	In the general sequential switching schemes, the LVRT can be divided into four stages: pre-fault (stage 1), during-fault (stage 2), early post-fault (stage 3), and late post-fault (stage 4) \cite{Ref_GB}. The DFIG has to implement corresponding controls at each stage to meet different requirements. In stages 1 and 4, normal control is employed to ensure stability \cite{Ref_10}. In stage 2, the DFIG needs to switch to the LVRT control to quickly support terminal voltage \cite{Ref_11}. In stage 3, the ramp control is employed to limit the active power recovery speed \cite{Ref_GB}. Clearly the high-dimensional, nonlinear, and event-driving switch characteristics of the DFIG grid-tied system make it very  difficult to analyze.

	Under the during-fault stage 2, it is found that the hardware protection circuit and the AC current control dynamics can be ignored and the phase-lock loop (PLL) dynamics for synchronizing with the grid is important. Under this situation, the system can be simplified as a second-order system. If the system loses its equilibrium point during the fault, a phenomenon of loss of synchronism is reported \cite{Ref_12}. Due to its structural similarity with the swing equation of the SG, it is called generalized swing equation (GSE) \cite{Ref_15,Ref_16,Ref_17}. So far, some classical analytical methods including the Lyapunov method \cite{Ref_18}, equal area criterion (EAC) method \cite{Ref_19}, phase portrait approach \cite{Ref_Wang}, and perturbation method \cite{Ref_21}, etc. are developed. In addition, taking into account saturation nonlinearities, more complicated nonlinear dynamical behaviors are reported very recently \cite{Ref_22},\cite{Ref_23}. However, it should be notable that the during-fault stage is only a fraction of the whole LVRT processes, and strictly the system stability should be judged after all four stages.

	The active power recovery stage 3 in the LVRT is also important \cite{Ref_24},\cite{Ref_25}. So far, transient stability analysis including the dynamics of stage 3 mainly relies on electromagnetic transient (EMT) simulation. Theoretical analysis still lacks. Except these studies, after greatly simplifying the dynamics of LVRT, the influence of the DFIG on the rotor angle stability of SG is studied \cite{Ref_26},\cite{Ref_28}. In recent studies \cite{Ref_PJJ, Ref_GHH, Ref_ZYY2}, transient models considering the complete LVRT processes are constructed, and TSS analysis is no longer limited to stage 2. However, stage 3 exhibits non-autonomous characteristics \cite{Ref_GHH}, making it difficult to analyze its transient dynamic. Clearly detailed dynamical property of stage 3 and its impact on the transient stability remain to be studied.

	Therefore, this paper aims to provide a system-level physical picture of the DFIG grid-tied system by considering the complete LVRT processes under severe faults, analyze the dominant factors and physical mechanism for the TSS, and develop novel TSS assessment methods. The main contributions in terms of modeling, analysis, and assessment are as follows:
	
	1) In the transient modeling contribution, a transient reduced-order model is constructed for each stage, to uncover the dominant dynamical characteristics and clarify switching conditions.
	
	2) In the dynamics analysis contribution, the bulk dynamics in both stages 2 and 3 can be caught by the GSE, the initial state of stage 3 plays a decisive role, and the TSS can be completely determined by the condition if it is within or out of the BOA of stage 3.
	
	3) In the stability assessment contribution, two efficient BOA-based and EAC-based methods are proposed. The TSS can be assessed immediately at the initial moment of stage 3.

	The rest of this article is structured as follows. In Section II, the topology structure and sequential switching characteristics of the DFIG system are introduced. In Section III, the transient mechanism model is constructed for all four stages. In Section IV, an EAC-based assessment method for permanent voltage-dip faults is introduced, as most of previous researchers have studied. In Section V, by studying the dynamical characteristics in stage 3, the salient effect of its initial state is uncovered. In Section VI, an extended EAC assessment method is developed and widely verified by simulations. In the end, the conclusions of this paper are made.

	\section{DFIG System Considering LVRT}	
	
\begin{figure}[!t]
			\centering
			\includegraphics[width=0.95\linewidth]{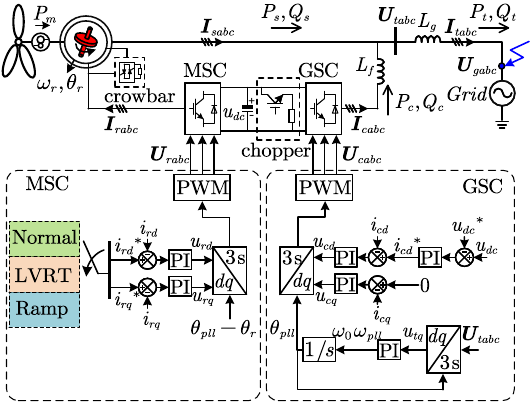}
			\caption{Schematic show of the DFIG system considering LVRT.}
			\label{topology}
		\end{figure}

	The topology structure and control scheme of the DFIG system are shown in \figurename \ref{topology}. Various types of energy storage element are included, such as the AC inductor, DC capacitor, and mechanical rotor, and the different dynamical responses of these elements are attributed to their own storage capacities. Correspondingly, the cascaded controllers adopt matched bandwidths to control these elements. For example, the bandwidth of the inner loop should be  designed about ten times larger than that of the outer loop.
	
	According to the grid code in China, when the  terminal voltage positive-sequence-component of DFIG $U_t$ is lower than 0.8p.u., the LVRT control needs to be switched on \cite{Ref_GB}. During the LVRT processes, the crowbar and chopper are triggered to protect the converter and capacitor. If the fault is small under the so-called shallow fault, the DFIG maintains the normal control, as shown in the solid line part of \figurename \ref{topology}. For the grid-side converter (GSC), it is composed of the DC-voltage control and the AC current control. The DC-voltage control aims to maintain the DC-voltage stable. The reactive power branch is to control the power factor, and usually the reference current $i_{c q}^{*}$ is set to zero \cite{Ref_Abad}. For the machine-side converter (MSC), it includes the maximum power point tracking control, pitch control, rotor speed control (RSC), terminal voltage control (TVC), and AC current control. Since the maximum power point tracking and the pitch controls exhibit a comparatively slow performance, they are ignored in this paper. The RSC aims to keep rotor speed stable, and the TVC aims to ensure the terminal voltage stable.

\begin{figure}[!t]
		\centering
				\includegraphics[width=0.7\linewidth]{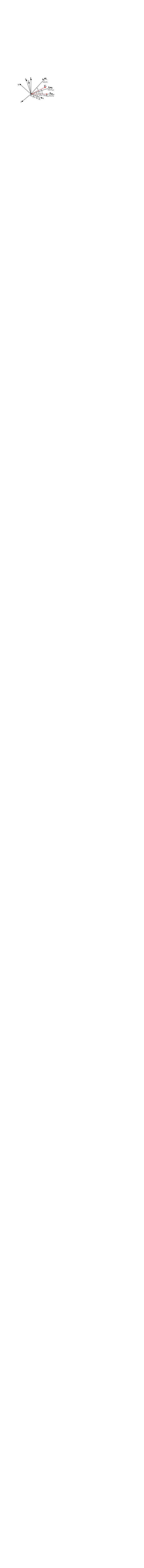}
		\caption{Schematic shows of three-phase stationary $abc$ reference frame, $xy$ reference frame, $dq$ common reference frame, and rotor reference frame, where $\omega_0$$\omega _{\rm \mathit{g}}$, $\omega_0$$\omega _{\rm \mathit{pll}}$, and $\omega_0$$\omega _{\rm \mathit{r}}$ represent the rotation angular frequencies of the $xy$, $dq$, and rotor reference frames, respectively. $\omega_0$ = 2$\pi$$f_0$.}
		\label{Vector}
	\end{figure}

	To unify coordinate of variables, the variables of the GSC and the MSC prompt the $dq$ reference frame provided by the PLL. \figurename \ref{Vector} shows the relationships between these different reference frames, where the phase mismatch of the rotating vector relative to the $xy$ reference frame is denoted by $\varphi$. For example, $\varphi _{\rm \mathit{pll}}=\theta _{\rm \mathit{pll}}-{\omega_0}{\omega _{\rm \mathit{g}}}t$. For the fundamental angular frequency, $\omega_0 = 2{\pi}{f_0}$, and for the grid angular frequency, $\omega _{\rm \mathit{g}}=1$ p.u..

	\begin{figure}[!t]
		\centering
		\includegraphics[width=1.0\linewidth]{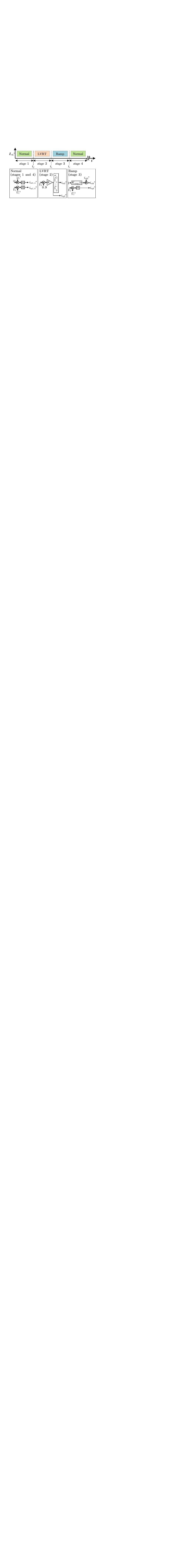}
		\caption{Schematic shows of typical switching controls under different stages. }
		\label{stage}
	\end{figure}
	
	Under a severe fault, the LVRT should be switched on according to the sequential switching schemes, as shown in the dashed line part of \figurename \ref{topology}. In particular, the outer loop control of the MSC should vary under different stages, as shown in \figurename \ref{stage}. The control and switching conditions for each stage will be studied in detail afterwards. Here $t_f$, $t_c$, and $t_r$ denote the times for the fault occurrence, clearing, and  ramp-ending, respectively. Throughout this paper, the subscripts 1, 2, 3, and 4 are used to denote the corresponding stages.

\section{Transient Model considering LVRT}
To catch the dominant dynamics and focus on the core factor of the TSS, the following assumptions are made:
	
	1) The fast dynamics of the inductance, including the line inductance $L_{\rm \mathit{g}}$, stator inductance $L_{\rm \mathit{s}}$, and rotor inductance $L_{\rm \mathit{r}}$, are neglected. Given that the response speed of the AC current control is very fast, its dynamics is also ignored, and thus  ${i\rm_{\mathit{rd}}}={i\rm_{\mathit{rd}}}^{*}$.
	
	2) According to the grid code of China, it is mandatory to switch into the LVRT control within 60 ms under severe faults \cite{Ref_GB}. Therefore, the operating time of the crowbar and demagnetization control is extremely brief and can be ignored. Additionally, the overheating limit of the chopper is neglected.
	
	3) The GSC is considered as a controlled current source and its dynamics is ignored. Hence the relation between the stator current $i\rm_{\mathit{sd}}$ and the output current $i\rm_{\mathit{cd}}$ of the GSC is simple: $i_{\rm \mathit{cd}}$ = ($\omega_{\rm \mathit{r}} -1$) $i_{\rm \mathit{sd}}$.

	\subsection{Stage 1: $pre$-$fault$}

			In stage 1, the DFIG adopts the normal control. For the MSC, the typical proportional-integral (PI) control is adopted in the RSC and TVC, whose differential equations are
			\begin{equation}
				\label{eq_RSC}
				\left\{ \begin{array}{l}
					{\dot i }_{{\rm{\mathit{rd}1}}} \; = {k_{{\rm{\mathit{pw}}}}} { {\dot \omega_ {\rm{\mathit{r}1}}} }   + {k_{{\rm{\mathit{iw}}}}} ({ {\omega_ {\rm{\mathit{r}1}}} }-{ {{\omega_r}^{*}} })\\
					{\dot i }_{{\rm{\mathit{rq}1}}} \; = {k_{{\rm{\mathit{pV}}}}} { {\dot {U}_ {\rm{\mathit{t}1}}} }   + {k_{{\rm{\mathit{iV}}}}} ({ {{U}_ {\rm{\mathit{t}1}}} } -{ {{U_t}^{*}} })
				\end{array} \right.
			\end{equation}
		where $k\rm_{\mathit{pw}}$ and $k\rm_{\mathit{iw}}$ are the proportional and integral coefficients of the RSC, respectively, $k\rm_{\mathit{pV}}$ and $k\rm_{\mathit{iV}}$ are the proportional and integral coefficients of the TVC, respectively, and ${\omega_r}^{*}$ and ${U_t}^{*}$ are the reference values of the rotor speed and the terminal voltage, respectively.
		
		The motion equation of the rotor is  
			\begin{equation}
				{\dot \omega }_{{\rm{\mathit{r}1}}} 
				\; = { {P{\rm_{\mathit{in}}}-P\rm_{\mathit{t}1} } \over {2\mathit{H}{\omega_{\rm{\mathit{r}1}}}}}
				\label{eq_Rotor}
			\end{equation}
			where $H$ represents the inertial time constant. $P\rm_{\mathit{in}}$ and $P\rm_{\mathit{t}1}$ represent the mechanical input power and the electromagnetic output power, respectively.

			For the PLL dynamics, the relation between $\omega\rm_{\mathit{pll}1}$ and the integrator output $x\rm_{\mathit{pll}1}$ is ${\omega_0}{\omega\rm_{\mathit{pll}1}}$ = ${\omega_0}{x\rm_{\mathit{pll}1}} + k\rm_{\mathit{ppll}}$$u\rm_{\mathit{tq}1}$, and the corresponding differential equations are
			\begin{equation}
				\label{eq_PLL}
				\left\{ \begin{array}{l}
					{\dot x }_{{\rm{\mathit{pll}1}}} \; = ({ k\rm_{\mathit{ipll}} }  {u_{\rm{\mathit{tq}1}}})/{\omega_0}\\
					{\dot \varphi }_{{\rm{\mathit{pll}1}}} \; = {\omega_0} (({ k\rm_{\mathit{ppll}} }  {u_{\rm{\mathit{tq}1}}})/{\omega_0}
					+ {x}_{{\rm{\mathit{pll}1}}} - 1) 
				\end{array} \right.
			\end{equation}
			where $k\rm_{\mathit{ppll}}$ and $k\rm_{\mathit{ipll}}$ are the proportional and integral coefficients of the PLL, respectively.

			The total output currents of the DFIG, $\bm{I}_{\rm\mathit{t}1}$, in the $dq$-axis are
			 	\begin{equation}
			 				\label{eq_itdq}
			 				\left\{ \begin{array}{l}
			 					{i}_{{\rm{\mathit{td}1}}} \; = {i}_{{\rm{\mathit{sd}1}}} + {i}_{{\rm{\mathit{cd}1}}} ={\omega\rm_{\mathit{r}1}}{i}_{{\rm{\mathit{sd}1}}} \\
			 					{i}_{{\rm{\mathit{tq}1}}} \; = {i}_{{\rm{\mathit{sq}1}}} + {i}_{{\rm{\mathit{cq}1}}} = {i}_{{\rm{\mathit{sq}1}}}
			 				\end{array} \right.
			 			\end{equation}
			As the transmission line dynamics is ignored, the static inductance connecting the terminal voltage $\bm{U}_{\rm\mathit{t}1}$ and the terminal current $\bm{I}_{\rm\mathit{t}1}$ is described by
			\begin{equation}
							\label{eq_Xg}
							\left\{ \begin{array}{l}
								{u}{{\rm_{\mathit{td}1}}} \; = {U}_{\rm{\mathit{g}1}}\cos\varphi_{\rm_{\mathit{pll}1}} - {X_g}{i}_{{\rm{\mathit{tq}1}}} \\
								{u}{{\rm_{\mathit{tq}1}}} \; = -{U}_{\rm{\mathit{g}1}}\sin\varphi_{\rm_{\mathit{pll}1}} + {X_g}{i}_{{\rm{\mathit{td}1}}} 
							\end{array} \right.
			\end{equation}

			For the asynchronous machine in the DFIG, its resistance is negligible. By neglecting the rapid dynamics of the flux linkage, the stator flux equation for $\bm{U}_{\rm\mathit{t}1}$ and the stator and rotor currents, $\bm{I}_{\rm\mathit{s}1}$ and $\bm{I}_{\rm\mathit{r}1}$, becomes \cite{Ref_Wu}
			\begin{equation}
				{j}{X_\mathit{m}}\bm{I}_{\rm\mathit{r}1}    \; =\bm{U}_{\rm\mathit{t}1}+  {j}{X_\mathit{s}}\bm{I}_{\rm\mathit{s}1}
				\label{eq_linkage}
			\end{equation}
			or, equivalently in the $dq$-axis
			\begin{equation}
							\label{eq_isdq}
							\left\{ \begin{array}{l}
								{i}{{\rm_{\mathit{sd}1}}} \; = ({X_m}{i}{{\rm_{\mathit{rd}1}}} - {u}_{\rm\mathit{tq}1}) / {X_s}  \\
								{i}{{\rm_{\mathit{sq}1}}} \; = ({X_m}{i}{{\rm_{\mathit{rq}1}}} + {u}_{\rm\mathit{td}1}) / {X_s} 
							\end{array} \right.
			\end{equation}
			where $X_\mathit{m}$ and $X_\mathit{s}$ are the per-unit values of the mutual reactance and the stator reactance, respectively.

			Combing (\ref{eq_itdq}), (\ref{eq_Xg}), (\ref{eq_isdq}) and eliminating the variables of the stator current, an explicit relation between $\bm{U}_{\rm\mathit{t}1}$ and $\bm{I}_{\rm\mathit{r}1}$ is 
			\begin{equation}
				\label{eq_utsimp}
				\left\{ \begin{array}{l}
					{u}{{\rm_{\mathit{td}1}}} \; = {a}{U}_{\rm{\mathit{g}1}}\cos\varphi_{\rm_{\mathit{pll}1}} - {b}{{X}_\mathit{g}}{i_{\rm{\mathit{rq}1}}} \\
					{u}{{\rm_{\mathit{tq}1}}} \; = -{c}{U}_{\rm{\mathit{g}1}}\sin\varphi_{\rm_{\mathit{pll}1}} + {d}{{X}_\mathit{g}}{i_{\rm{\mathit{rd}1}}}
				\end{array} \right.
			\end{equation}
			where the correction coefficients ($a$, $b$, $c$, and $d$) are
						\begin{equation}
							\label{eq_Coeff}
							\left\{ \begin{array}{l}
								{a} \; = {X_\mathit{s}}/({X_\mathit{s} + X_\mathit{g}}) \\
								{b} \; = {X_\mathit{m}}/({X_\mathit{s} + X_\mathit{g}}) \\
								{c} \; = {X_\mathit{s}}/({X_\mathit{s} + {\omega\rm_{\mathit{r}1}}X_\mathit{g}}) \\
								{d} \; =  {\omega\rm_{\mathit{r}1}}{X_\mathit{m}}/({X_\mathit{s} +  {\omega\rm_{\mathit{r}1}}X_\mathit{g}})
							\end{array} \right.
						\end{equation}
            At the stable equilibrium point, $a=0.89$, $b=0.85$, $c=0.87$, and $d=1$. Clearly here $\bm{U}_{\rm\mathit{t}1}$ and $\bm{I}_{\rm\mathit{r}1}$ show the only difference of these scaled, correction coefficients, to be compared with the network equations in (\ref{eq_Xg}).

			Finally, the differential-algebraic equations (DAEs) in stage 1 for the major dynamics of the RSC, TVC, PLL, and rotor are
			\begin{equation}
				\label{eq_stage1_OE}
				\left\{ \begin{array}{l}
					{\dot \omega }_{{\rm{\mathit{r}1}}} \; = ({ P\rm_{\mathit{in}} } - { P\rm_{\mathit{t}1} }) / ({2\mathit{H}{\omega_{\rm{\mathit{r}1}}}})\\	[1mm]
					{\dot i }_{{\rm{\mathit{rd}1}}} \; = {k_{{\rm{\mathit{pw}}}}} { {\dot \omega_ {\rm{\mathit{r}1}}} }   + {k_{{\rm{\mathit{iw}}}}} ({ {\omega_ {\rm{\mathit{r}1}}} }-{ {{\omega_r}^{*}} })\\[1mm]
					{\dot i }_{{\rm{\mathit{rq}1}}} \; = {k_{{\rm{\mathit{pV}}}}} { {\dot {U}_ {\rm{\mathit{t}1}}} }   + {k_{{\rm{\mathit{iV}}}}} ({ {{U}_ {\rm{\mathit{t}1}}} } -{ {{U_t}^{*}} }) \\[1mm]
					{\dot x }_{{\rm{\mathit{pll}1}}} \; = ({ k\rm_{\mathit{ipll}} }  {u_{\rm{\mathit{tq}1}}})/{\omega_0}\\[1mm]
					{\dot \varphi }_{{\rm{\mathit{pll}1}}} \; = {\omega_0} (({ k\rm_{\mathit{ppll}} }  {u_{\rm{\mathit{tq}1}}})/{\omega_0}
					+ {x}_{{\rm{\mathit{pll}1}}} - 1) 
				\end{array} \right.
			\end{equation}
			\begin{equation}
				\label{eq_stage1_DE}
				\left\{ \begin{array}{l}
					{u}{{\rm_{\mathit{td}1}}} \; = {a}{U}_{\rm{\mathit{g}1}}\cos\varphi_{\rm_{\mathit{pll}1}} - {b}{{X}_\mathit{g}}{i_{\rm{\mathit{rq}1}}} \\[1mm]
					{u}{{\rm_{\mathit{tq}1}}} \; = -{c}{U}_{\rm{\mathit{g}1}}\sin\varphi_{\rm_{\mathit{pll}1}} + {d}{{X}_\mathit{g}}{i_{\rm{\mathit{rd}1}}}\\[1mm]
					{i}{{\rm_{\mathit{td}1}}} \; = {\omega\rm_{\mathit{r}1}}({X_m}{i}{{\rm_{\mathit{rd}1}}} - {u}_{\rm\mathit{tq}1}) / {X_s} \\[1mm]
					{i}{{\rm_{\mathit{tq}1}}} \; = ({X_m}{i}{{\rm_{\mathit{rq}1}}} + {u}_{\rm\mathit{td}1}) / {X_s}\\[1mm] 
					{P}_{\rm{\mathit{t}1}} \;= {u}{{\rm_{\mathit{td}1}}}{i}{{\rm_{\mathit{td}1}}} + 
										{u}{{\rm_{\mathit{tq}1}}} {i}{{\rm_{\mathit{tq}1}}}\\[1mm]
					{U}{{\rm_{\mathit{t}1}}} \; = \sqrt{{u_{\rm{\mathit{td}1}}}^{2} + {u_{\rm{\mathit{tq}1}}}^{2}}
				\end{array}\vspace{2ex} \right.
			\end{equation}

Obviously, there exist a stable equilibrium point (SEP), ${\bf{x}}_s$:
			\begin{equation}
				\label{eq_SEP1}
				\left\{ 
				{\begin{array}{l}
					{ \omega }_{{\rm{\mathit{r}1,\mathit{s}}}} \; = {\omega\rm_{\mathit{r}}}^{*}\\	[2mm]
					{i }_{{\rm{\mathit{rd}1,\mathit{s}}}} \; = {\dfrac{ {X\rm_{\mathit{s}}} {P\rm_{\mathit{in}}}   }     {  {{X\rm_{\mathit{m}}} } {\omega\rm_{\mathit{r}}}^{*} }}\\[2mm]
					{i }_{{\rm{\mathit{rq}1,\mathit{s}}}} \; = \dfrac{ {X_s}{U\rm_{\mathit{g}1}}{\cos{\varphi\rm_{\mathit{pll}1,\mathit{s}}}} -{X_s}{X_g}{{U_t}^{*}} }{{X_g}{X_m}} \\[2mm]
					{x }_{{\rm{\mathit{pll}1,\mathit{s}}}} \; = 1\\[2mm]
					{\varphi }_{{\rm{\mathit{pll}1,\mathit{s}}}} \; = {\arcsin({\dfrac{ {P\rm_{\mathit{in}}}  {X\rm_{\mathit{g}}} }     { {U\rm_{\mathit{g}1}} {{U\rm_{\mathit{t}}}^{*} }  }})}
				\end{array}\vspace{2ex}}\right.
			\end{equation}		
			and an unstable equilibrium point (UEP), ${\bf{x}}_u$:
						\begin{equation}
							\label{eq_UEP1}
							\left\{ 
							{\begin{array}{l}
								{ \omega }_{{\rm{\mathit{r}1,\mathit{u}}}} \; = {\omega\rm_{\mathit{r}}}^{*}\\	[2mm]
								{i }_{{\rm{\mathit{rd}1,\mathit{u}}}} \; = {\dfrac{ {X\rm_{\mathit{s}}} {P\rm_{\mathit{in}}}   }     {  {{X\rm_{\mathit{m}}} } {\omega\rm_{\mathit{r}}}^{*} }}\\[2mm]
								{i }_{{\rm{\mathit{rq}1,\mathit{u}}}} \; = \dfrac{ {X_s}{U\rm_{\mathit{g}1}}{\cos{\varphi\rm_{\mathit{pll}1,\mathit{u}}}} -{X_s}{X_g}{{U_t}^{*}} }{{X_g}{X_m}} \\[2mm]
								{x }_{{\rm{\mathit{pll}1,\mathit{u}}}} \; = 1\\[2mm]
								{\varphi }_{{\rm{\mathit{pll}1,\mathit{u}}}} \; = \pi-{\arcsin({\dfrac{ {P\rm_{\mathit{in}}}  {X\rm_{\mathit{g}}} }     { {U\rm_{\mathit{g}1}} {{U\rm_{\mathit{t}}}^{*} }  }})}
							\end{array}\vspace{2ex}}\right.
						\end{equation}
			
			Under the typical parameters in Appendix, ${\bf{x}}_s=[			
			1.2,0.7,-0.43,1,0.41]^T$ p.u. and ${\bf{x}}_u=[
			1.2,0.7,-4.25,1,2.73]^T$ p.u., where the superscript $T$ denotes transposition. Clearly ${\bf{x}}_s$ provides an initial stable operating point for the sequential stages.

	\subsection{Stage 2: $during$-$fault$} 
	When a severe fault occurs, e.g., $U_g$ dips from $1.0$ p.u. to a much smaller $U_{g2}$, the terminal voltage ${U\rm_{\mathit{t}2}}$ dips below 0.8 p.u. accordingly, causing the normal control to freeze. Stage 2 begins. Now the $dq$ current of the MSC is provided by the LVRT control. To rapidly support ${U\rm_{\mathit{t}2}}$, the reactive current ${i\rm_{\mathit{rq}2}}$ is injected in proportion to the magnitude of ${U\rm_{\mathit{t}2}}$. In addition, the active current ${i\rm_{\mathit{rd}2}}$ should be limited by the capacity of converter ${I}_{\rm{\mathit{max}}}$;
	\begin{equation}
		\label{eq_LVRT}
		\left\{ \begin{array}{l}
			{i}{{\rm_{\mathit{rq}2}}} \; = K_\mathit{e}(0.9 - U_{\rm{\mathit{t}}2})
			+{i}{{\rm_{\mathit{rq}1,\mathit{s}}}}  \\
			0  \leq {i}{{\rm_{\mathit{rd}2}}} \leq \sqrt{{I_{\rm{\mathit{max}}}}^{2} -{i_{\rm{\mathit{rq}2}}}^{2} } 
		\end{array} \right.
	\end{equation}
	where $i\rm_{\mathit{rq}1,\mathit{s}}$ is the initial reactive current of stage 1, and ${K}_e$ is the reactive current coefficient. As 1.5 $\leq$ ${K}_e$ $\leq$ 3 is often chosen in engineering, ${K}_e = 1.5$ is fixed in this paper.  ${I}_{\rm{\mathit{max}}}=1.1$ p.u.. 
	On the other hand, as $U\rm_{\mathit{t}2}$ in stage 2 only changes slightly, $i\rm_{\mathit{rq}2}$ in (\ref{eq_LVRT})	
	is often chosen as fixed by the initial value of $U\rm_{\mathit{t}2}$ at stage 2. In a contrast, $i\rm_{\mathit{rd}2}$ can be treated as an adjustable parameter, subjected by the capacity constraint determined by $i\rm_{\mathit{rq}2}$ and $I\rm_{\mathit{max}}$. 
	
	Now the algebraic equations in (\ref{eq_utsimp}) are unchanged, with the only one parameter change from $U\rm_{\mathit{g}1}$ ($U\rm_{\mathit{g}1}$=1 p.u.) to $U\rm_{\mathit{g}2}$:
				\begin{equation}
					\label{eq_utsimp_s2}
					\left\{ \begin{array}{l}
						{u}{{\rm_{\mathit{td}2}}} \; = {a}{U}_{\rm{\mathit{g}2}}\cos\varphi_{\rm_{\mathit{pll}2}} - {b}{{X}_\mathit{g}}{i_{\rm{\mathit{rq}2}}} \\
						{u}{{\rm_{\mathit{tq}2}}} \; = -{c}{U}_{\rm{\mathit{g}2}}\sin\varphi_{\rm_{\mathit{pll}2}} + {d}{{X}_\mathit{g}}{i_{\rm{\mathit{rd}2}}}\\
					\end{array} \right.
				\end{equation}
Compared to the change of the rotor speed $\omega_r$, the duration times of stage 2 and the following stage 3 are short, and $c$ and $d$ can be regarded as constants. 
	
	Considering the PLL's dynamics which is determined by ${u_{\rm{\mathit{tq}2}}}$ only, and combing (\ref{eq_LVRT}), (\ref{eq_utsimp_s2}), the DAEs accompanying with the capacity constraint are 
	\begin{equation}
		\label{eq_stage2_OE}
		\left\{ \begin{array}{l}
					{\dot x }_{{\rm{\mathit{pll}2}}} \; = ({ k\rm_{\mathit{ipll}} }  {u_{\rm{\mathit{tq}2}}})/{\omega_0}\\
					{\dot \varphi }_{{\rm{\mathit{pll}2}}} \; = {\omega_0} (({ k\rm_{\mathit{ppll}} }  {u_{\rm{\mathit{tq}2}}})/{\omega_0}
										+ {x}_{{\rm{\mathit{pll}2}}} - 1) 
		\end{array} \right.
	\end{equation}

		\begin{equation}
	\label{eq_stage2_DE}
	\left\{ \begin{array}{l}		
	{u}{{\rm_{\mathit{tq}2}}} \; = -{c}{U}_{\rm{\mathit{g}2}}\sin\varphi_{\rm_{\mathit{pll}2}} + {d}{{X}_\mathit{g}}{i_{\rm{\mathit{rd}2}}}\\[1mm]
	0  \leq {i}{{\rm_{\mathit{rd}2}}} \leq \sqrt{{I_{\rm{\mathit{max}}}}^{2} -{i_{\rm{\mathit{rq}2}}}^{2} } \\
	\end{array} \right.
	\end{equation}


	\subsection{Stage 3: $early$ $post$-$fault$} 
 	When the fault is cleared and $U_g$ recovers, $U\rm_{\mathit{t}3}$ is restored instantly and the TVC is unfrozen. Stage 3 starts. However, to protect device, the active power needs to recover gradually. In this stage, the linear recovery of  $i_{\rm{\mathit{rd}3}}$ is constrained by the ramping rate $K_{\rm{\mathit{ramp}}}$. The differential equations of the outer loop control in the MSC are
	\begin{equation}
		\label{eq_Ramp}
		\left\{ \begin{array}{l}
			{\dot i }_{{\rm{\mathit{rd}3}}} \; = K_{\rm{\mathit{ramp}}}\\
			{\dot i }_{{\rm{\mathit{rq}3}}} \; = {k_{{\rm{\mathit{pV}}}}} { {\dot {U}_ {\rm{\mathit{t}3}}} }   + {k_{{\rm{\mathit{iV}}}}} ({ {{U}_ {\rm{\mathit{t}3}}} } -{ {{U_t}^{*}} }) 
		\end{array} \right.
	\end{equation}
    If the active power is restored too quickly, 
	the electrical variables of the DFIG will oscillate violently and 
	the unbalanced power on the shaft will be intensified, probably leading to torsional vibration. To suppress these effects, the ramping rate is generally set as relatively small \cite{Ref_24},\cite{Ref_26}. This is very different from the other renewable devices, such as the permanent magnet synchronous generator (PMSG) or photovoltaic (PV) systems. Oppositely, it also cannot be chosen too small, which might cause frequency issues. Therefore, $K_{\rm{\mathit{ramp}}} > 0.2$ is often chosen in engineering \cite{Ref_GB}. In this paper, $K_{\rm{\mathit{ramp}}}$ = 0.8 is generally selected. 
	
	Now with the major controls including the outer loop controls of the MSC and the PLL, the DAEs in stage 3 are
	\begin{equation}
		\label{eq_stage3_OE}
		\left\{ \begin{array}{l}
			{\dot i }_{{\rm{\mathit{rd}3}}} \; = K_{\rm{\mathit{ramp}}}\\
			{\dot i }_{{\rm{\mathit{rq}3}}} \; = {k_{{\rm{\mathit{pV}}}}} { {\dot {U}_ {\rm{\mathit{t}3}}} }   + {k_{{\rm{\mathit{iV}}}}} ({ {{U}_ {\rm{\mathit{t}3}}} } -{ {{U_t}^{*}} })  \\
			{\dot x }_{{\rm{\mathit{pll}3}}} \; = ({ k\rm_{\mathit{ipll}} }  {u_{\rm{\mathit{tq}3}}})/{\omega_0}\\[1mm]
			{\dot \varphi }_{{\rm{\mathit{pll}3}}} \; = {\omega_0} (({ k\rm_{\mathit{ppll}} }  {u_{\rm{\mathit{tq}3}}})/{\omega_0}
								+ {x}_{{\rm{\mathit{pll}3}}} - 1) 
		\end{array} \right.
	\end{equation}
	\begin{equation}
		\label{eq_stage3_DE}
		\left\{ \begin{array}{l}
			
			{u}{{\rm_{\mathit{td}3}}} \; = {a}{U}_{\rm{\mathit{g}3}}\cos\varphi_{\rm_{\mathit{pll}3}} - {b}{{X}_\mathit{g}}{i_{\rm{\mathit{rq}3}}} \\[1mm]					
			{u}{{\rm_{\mathit{tq}3}}} \; = -{c}{U}_{\rm{\mathit{g}3}}\sin\varphi_{\rm_{\mathit{pll}3}} + {d}{{X}_\mathit{g}}{i_{\rm{\mathit{rd}3}}}\\[1mm]
			{U}{{\rm_{\mathit{t}3}}} \; = \sqrt{{u_{\rm{\mathit{td}3}}}^{2} + {u_{\rm{\mathit{tq}3}}}^{2}}\\[1mm]
		\end{array} \right.
	\end{equation}
	
\subsection{Stage 4: $late$ $post$-$fault$} 
	When the active current arrives at the initial level, i.e., ${i\rm_{\mathit{rd}4}}= {i\rm_{\mathit{rd}1,\mathit{s}}}$, the normal control is fully restored. Therefore, the model in stage 4 becomes exactly the same as that in stage 1.

	\subsection{Simulation verification and summary} 
\begin{figure}[!t]
					\centering
					\includegraphics[width=1.0\linewidth]{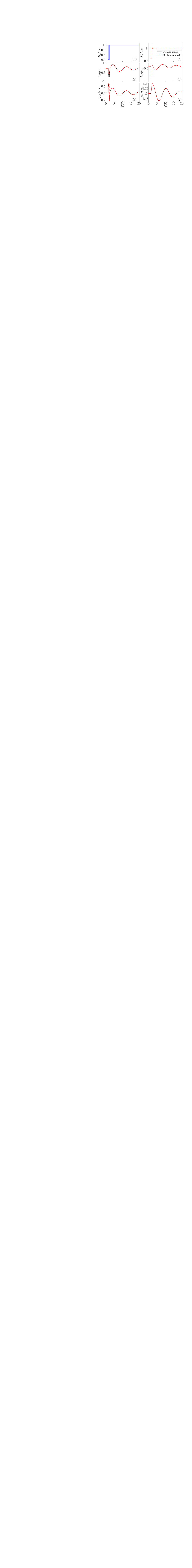}
					\caption{(a)-(f) Plots of $U_g$, $U_t$, $i_{\rm\mathit{rd}}$, $i_{\rm\mathit{rq}}$, $\varphi_{{\rm{\mathit{pll}}}}$, and $\omega_r$, respectively, for comparison of the mechanism model with the detailed EMT model.}
					\label{Sim_Compare}
				\end{figure}

	The above mechanism model containing the whole four stages has been broadly verified in MATLAB/Simulink by comparison with the detailed EMT model. The parameters are listed in Appendix. As one example, at $t_f$ = 1.5 s, $U_g$ dips from 1 p.u. to 0.4 p.u.. At $t_c$ = 2.1 s, the fault is cleared. The fault duration time $t_c$ – $t_f$ = 0.6 s. During the fault,  ${i\rm_{\mathit{rd}2}}=0.34$ p.u. is chosen.	
	The comparative results are shown in \figurename \ref{Sim_Compare}, where clearly demonstrates that the mechanism model is consistent with the detailed model, although some discernible fast dynamics at the switching moments of each stage are missed. All these are understandable.

		In summary, stage 1 provides an initial stable operating point. When a severe fault occurs and $U_t$ drops below 0.8 p.u., the LVRT control is activated and stage 2 starts. Stage 2 (under $U_{g2}$) provides voltage support and the active-power current $i\rm_{\mathit{rd}2}$ can be regarded as an adjustable parameter. After the fault is cleared, $U_t$ recovers instantly. Stage 3 (under $U_{g3}$) starts and  $i\rm_{\mathit{rd}3}$ needs to recover gradually. When the $i\rm_{\mathit{rd}3}$ recovers to the initial level, stage 4 starts. Therefore, the whole LVRT processes of the DFIG system involve four stages, and basically the TSS should be determined by whether the final stage 4 can settle in a SEP of stage 4 \cite{Ref_Ziqian}. Afterwards, it will be very interesting to see that based on the dynamical characteristics of stage 3, this criterion can be greatly loosed and the TSS can become easier.

\section{TSS analysis of stage 2}
\label{sect_stage 2}				
		There are many studies focusing on the TSS of stage 2 under the condition of permanent faults or the concept of so-called device stability \cite{Ref_18,Ref_19,Ref_Wang,Ref_21}, namely, only if the grid-tied device keeps synchronization on stage 2, the system can be stable and the following stages 3 and 4 can be completely ignored. It is necessary to start from this simple case first. 
		
\subsection{Generalized swing equation}
						According to the DAEs in stage 2 in (\ref{eq_stage2_OE}) and (\ref{eq_stage2_DE}), the following GSE with the pure ordinary differential equation can be derived,
						\begin{equation}
								{M_{\rm{\mathit{eq}2}}}{\ddot \varphi }_{{\rm{\mathit{pll}2}}} \; = {P_{\rm{\mathit{m}2}}} - {P_{\mathit{e}2}} - {D_{\rm{\mathit{eq}2}}}{\dot \varphi }_{{\rm{\mathit{pll}2}}}
							\label{eq_GSE}
						\end{equation}
						where
						\begin{equation}
							\label{eq_stage3_GSE}
							\left\{ \begin{array}{l}
								{P_{\rm{\mathit{m}2}}} \; = {d}{X_{\mathit{g}}}{i_{{\rm{\mathit{rd}2}}}} \\[1mm]
								{P_{\rm{\mathit{e}2}}} \; = {c}{U{\rm_{\mathit{g}2}}}\sin{\varphi_{{\rm{\mathit{pll}2}}}}\\[1mm]
								{M_{\rm{\mathit{eq}2}}} \; = \dfrac{1}{k_{\rm\mathit{ipll}}}\\[1mm]
								{D_{\rm{\mathit{eq}2}}} \; = {c}\dfrac{k_{\rm\mathit{ppll}}}{k_{\rm\mathit{ipll}}}{U_{\rm{\mathit{g}2}}}\cos{\varphi_{\rm\mathit{pll}2}}
								
							\end{array} \right.
						\end{equation}
						Here ${P_{\rm{\mathit{m}2}}}$, ${P_{\rm{\mathit{e}2}}}$, ${M_{\rm{\mathit{eq}2}}}$, and ${D_{\rm{\mathit{eq}2}}}$ represent the equivalent mechanical power, electromagnetic power, inertia, and damping, respectively. Clearly, $P\rm_{\mathit{m}2}$ is a constant, depending on $i\rm_{\mathit{rd}2}$, and $P\rm_{\mathit{e}2}$ is a sinusoidal function of ${\varphi_{{\rm{\mathit{pll}2}}}}$, depending on $U\rm_{\mathit{g}2}$. Different from that 
						the PMSG and PV control the current of the GSC, the 			
						DFIG usually controls the current of the MSC to achieve LVRT. 
						This difference gives rise to the correction coefficients $c$ and $d$ in the GSE.

\subsection{EAC-based TSS assessment for voltage-dip permanent faults}

\begin{figure}[!t]
		\centering
		\includegraphics[width=.8 \linewidth]{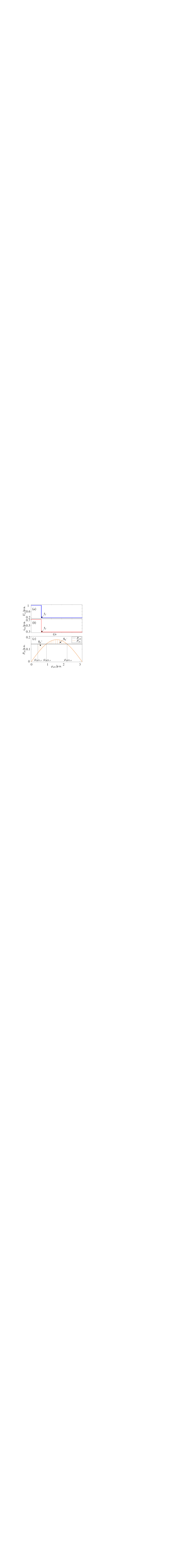}
		\caption{(a) and (b) Plots of $U_g$ and $i\rm_{\mathit{rd}}$, and (c) EAC-based TSS analysis for permanent faults.}
		\label{EAC_stage2}
	\end{figure}

				After ignoring the damping term in the GSE, the EAC can be used to analyze the TSS in stage 2 for permanent faults, as shown in \figurename \ref{EAC_stage2}. Before the fault, the system is working at the initial operating point of stage 1, as shown in (\ref{eq_SEP1}),
						\begin{equation}
								{\varphi\rm_{\mathit{pll}1,\mathit{s}}} ={\arcsin({\dfrac{ {P\rm_{\mathit{in}}}  {X\rm_{\mathit{g}}} }     { {U\rm_{\mathit{g}1}} {{U\rm_{\mathit{t}}}^{*} }  }})}=\arcsin(\dfrac{d{X_g}{i_{\mathit{rd}1}}}{c{U_{\mathit{g}1}}})
							\label{eq_SEPf1}
						\end{equation}
				The second equality comes from the second equation in (\ref{eq_utsimp}) under the steady-state value ${u}{{\rm_{\mathit{tq}2,s}}}=0$.
	
				Without losing generality, when $U_g$ dips (e.g., $U_g=$0.2 p.u.) and a permanent fault occurs at $t_f$, stage 2 starts. $i\rm_{\mathit{rq}2}= -0.93$ p.u. and $i\rm_{\mathit{rd}2} = 0.28$ p.u. are chosen. The equivalent power angle (sinusoidal curve) $P\rm_{\mathit{m}2}$ and the equivalent constant mechanical power (horizontal line) $P\rm_{\mathit{e}2}$ are shown in  \figurename \ref{EAC_stage2}(c). The SEP and UEP of stage 2 are
					\begin{equation}
						\label{eq_SEP2}
						\left\{ \begin{array}{l}
							{{\varphi}{{\rm_{\mathit{pll}2,\mathit{s}}}}} \; = \arcsin(\dfrac{d{X_g}{i_{\mathit{rd}2}}}{c{U_{\mathit{g}2}}})   \\[1mm]					
							{{\varphi}{{\rm_{\mathit{pll}2,\mathit{u}}}}} \; = {\pi}- {\varphi}{{\rm_{\mathit{pll}2,\mathit{s}}}}\\[1mm]
						\end{array} \right.
					\end{equation}
				The difference between ${{\varphi}{{\rm_{\mathit{pll}2,\mathit{s}}}}}$ and ${{\varphi}{{\rm_{\mathit{pll}1,\mathit{s}}}}}$ in (\ref{eq_SEPf1})
and (\ref{eq_SEP2})	lies in the different values of $U_g$ and $i\rm_{\mathit{rd}}$. When a fault occurs, since ${P\rm_{\mathit{m}2}} > {P\rm_{\mathit{e}2}}$ at ${\varphi\rm_{\mathit{pll}1,\mathit{s}}}$, ${\varphi\rm_{\mathit{pll}2}}$ will accelerate. When ${\varphi\rm_{\mathit{pll}2,\mathit{s}}}<{\varphi\rm_{\mathit{pll}2}}<{\varphi\rm_{\mathit{pll}2,\mathit{u}}}$, as ${P\rm_{\mathit{m}2}} < {P\rm_{\mathit{e}2}}$, ${\varphi\rm_{\mathit{pll}2}}$ will decelerate. This happens until it arrives at ${\varphi\rm_{\mathit{pll}2,\mathit{u}}}$. Therefore, the accelerating area  ${S_2}^{+}$ and the maximal decelerating area ${S_2}^{-}$ are
									\begin{equation}
										\label{eq_SAD}
										\left\{ \begin{array}{l}	
											{S_2}^{+} \; = \int_{{\varphi\rm_{\mathit{pll}1,\mathit{s}}}}^{{\varphi\rm_{\mathit{pll}2,\mathit{s}}}} ({P_{\rm{\mathit{m}2}}}-{P_{\rm{\mathit{e}2}}}) d{\varphi_{{\rm{\mathit{pll}2}}}}   \\[1mm]					
											{S_2}^{-} \; = \int_{{\varphi\rm_{\mathit{pll}2,\mathit{s}}}}^{{\varphi\rm_{\mathit{pll}2,\mathit{u}}}} ({P_{\rm{\mathit{e}2}}}-{P_{\rm{\mathit{m}2}}}) d{\varphi_{{\rm{\mathit{pll}2}}}} \\[1mm]
										\end{array} \right.
									\end{equation}

				Based on the EAC, to ensure the TSS of stage 2 under the permanent faults, the following condition needs to meet:
						\begin{equation}
								{S_2}^{+} \leq  {S_2}^{-}
							\label{eq_TSS2}
						\end{equation}

				According to (\ref{eq_stage3_GSE}), for a larger  $U\rm_{\mathit{g}2}$, $P\rm_{\mathit{e}2}$ becomes larger for a steeper curve, which is beneficial to the TSS. By a smaller $i\rm_{\mathit{rd}2}$, $P\rm_{\mathit{m}2}$ decreases for a lower horizontal line, and the TSS can also be improved. All these are in accordance with our common sense. Next the TSS considering the whole LVRT process will be studied and the impact of stage 3 dynamics will be concentrated.

\section{Transient analysis of stage 3}

 \subsection{Non-autonomous driving-response system}
 	Observing the DAEs in stage 3 in (\ref{eq_stage3_OE}) and (\ref{eq_stage3_DE}) carefully, one can find that actually they can be divided into the following two subsystems including the driving one:
 								\begin{equation}
 									\label{eq_stage3_3alpha_OE}
 									\left\{ \begin{array}{l}
 									{\dot x }_{{\rm{\mathit{pll}3}}} \; = ({ k\rm_{\mathit{ipll}} }  {u_{\rm{\mathit{tq}3}}})/{\omega_0}\\
 									{\dot \varphi }_{{\rm{\mathit{pll}3}}} \; = {\omega_0} (({ k\rm_{\mathit{ppll}} }  {u_{\rm{\mathit{tq}3}}})/{\omega_0}
 																+ {x}_{{\rm{\mathit{pll}3}}} - 1)  
 									\end{array} \right.
 								\end{equation}
 								\begin{equation}
 									\label{eq_stage3_3alpha_DE}
 									\left\{ \begin{array}{l}
 										{u}{{\rm_{\mathit{tq}3}}} \; = -{c}{U}_{\rm{\mathit{g}3}}\sin\varphi_{\rm_{\mathit{pll}3}} + {d}{{X}_\mathit{g}}{i_{\rm{\mathit{rd}3}}}\\[1mm]
 										{i }_{{\rm{\mathit{rd}3}}} \; = {i }_{{\rm{\mathit{rd}2}}} + K_{\rm{\mathit{ramp}}}(t-t_c)
 									\end{array} \right.
 								\end{equation}
and the other response one:
 								\begin{equation}
 									\label{eq_stage3_3beta_OE}
 										{\dot i }_{{\rm{\mathit{rq}3}}} \; = {k_{{\rm{\mathit{pV}}}}} { {\dot {U}_ {\rm{\mathit{t}3}}} }   + {k_{{\rm{\mathit{iV}}}}} ({ {{U}_ {\rm{\mathit{t}3}}} } -{ {{U}_ {\rm{\mathit{tref}}}} }) 
 								\end{equation}
 								\begin{equation}
 									\label{eq_stage3_3beta_DE}
 									\left\{ \begin{array}{l}
 										{u}{{\rm_{\mathit{td}3}}} \; = {a}{U}_{\rm{\mathit{g}3}}\cos\varphi_{\rm_{\mathit{pll}3}} - {b}{{X}_\mathit{g}}{i_{\rm{\mathit{rq}3}}} \\[1mm]					
 										{U}{{\rm_{\mathit{t}3}}} \; = \sqrt{{u_{\rm{\mathit{td}3}}}^{2} + {u_{\rm{\mathit{tq}3}}}^{2}}\\[1mm]
 									\end{array} \right.
 								\end{equation}
 						
 				Clearly in the driving subsystem, $u_{\rm\mathit{tq}3}$ is affected by $i_{\rm\mathit{rd}3}$, which depends on time. While for the response subsystem, $u_{\rm\mathit{td}3}$ is affected by  $\varphi_{\rm\mathit{pll}3}$ and $i_{\rm\mathit{rq}3}$, and $U_{\rm\mathit{t}3}$ is affected by $u_{\rm\mathit{tq}3}$ which should come from the driving subsystem.

 				In the driving subsystem, as $i_{\rm\mathit{rd}3}$ increases linearly, it exhibits the non-autonomous characteristics. However, if its dynamics changes slowly,  $i_{\rm\mathit{rd}3}$ can be approximately treated as a constant and hence the driving subsystem can be considered as a generalized autonomous system. Similar treatments have been widely used in the slow-fast non-autonomous analysis in mathematics \cite{Ref_34}. Therefore, the dominant dynamics of stage 3 can be viewed as a series of PLL second-order dynamics under a slow change of $i_{\rm\mathit{rd}3}$ and a fixed ${U}_{\rm{\mathit{g}3}}$. It is similar to the stage 2 dynamics essentially.

 		\subsection{Dynamical characteristics of stage 3}
 				
\begin{figure}[!t]
					\centering
					\includegraphics[width=.8 \linewidth]{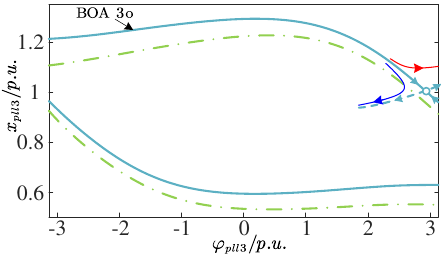} 
					\caption{Illustrations of different BOA boundaries under different values of $i_{\rm_{\mathit{rd}3}}$ and local dynamics near the UEP (open circle). The light blue solid (green dot-dashed) line represents the BOA at the initial (end) state of stage 3 for $i_{\rm{\mathit{rd}3}}=i_{\rm{\mathit{rd}2}}=0.34$ p.u. ($i_{\rm{\mathit{rd}3}}=i_{\rm{\mathit{rd}1,\mathit{s}}}=0.7$ p.u.)
					 }
					\label{red_blue}
					\end{figure}

 				With a constant $i\rm_{\mathit{rd}3}$, it can be similarly studied on the ${x\rm_{\mathit{pll}}}-\varphi\rm_{\mathit{pll}}$ plane. The different BOA boundaries under different $i_{\rm{\mathit{rd}3}}$'s are illustrated in \figurename \ref{red_blue}. The light blue solid (green dot-dashed) line represents the BOA at the initial (end) state of stage 3 for $i_{\rm{\mathit{rd}3}}=i_{\rm{\mathit{rd}2}}=0.34$ p.u. ($i_{\rm{\mathit{rd}3}}=i_{\rm{\mathit{rd}1,\mathit{s}}}=0.7$ p.u.) Since the initial state of stage 3 under $i\rm_{\mathit{rd}2}$ and $U_{g3}$				
 				will be the most concerned, its BOA  is emphasized by BOA 3o. Based on this comparison, it can be found that with a slow increase of $i_{\rm{\mathit{rd}3}}$, the BOA also moves to the lower-left part slowly and the bulk structure of the BOA is unchanged.

 				On the other hand, it is well known that the local dynamics near the UEP represented by an open circle always dominates the TSS. If the system is within the BOA 3o and near the unstable manifold of the UEP, it will quickly move away from the UEP. The moving speed is usually much faster than that of the BOA. On the contrary, if the system is out of the BOA 3o and on the other side of the unstable manifold of the UEP, it will quickly move away from the UEP to the upper-right direction. 
 				
 				Based on these two combined effects determined by the slow motion of the BOA and the local manifold structure near the UEP, it can easily derive that the initial state of stage 3 is dominant for the TSS, namely, if it is within the BOA 3o, the system will be stable, or otherwise, it will be unstable. 
 				Therefore, the following stages 3 and 4 after this particular initial moment of stage 3 do not need to be considered. Fundamentally different from the TSS for permanent faults in Section \ref{sect_stage 2}, here the LVRT effect is taken into account. All these analyses need to be verified by simulations.

\begin{figure}[!t]
	\centering
	\includegraphics[width=0.85\linewidth]{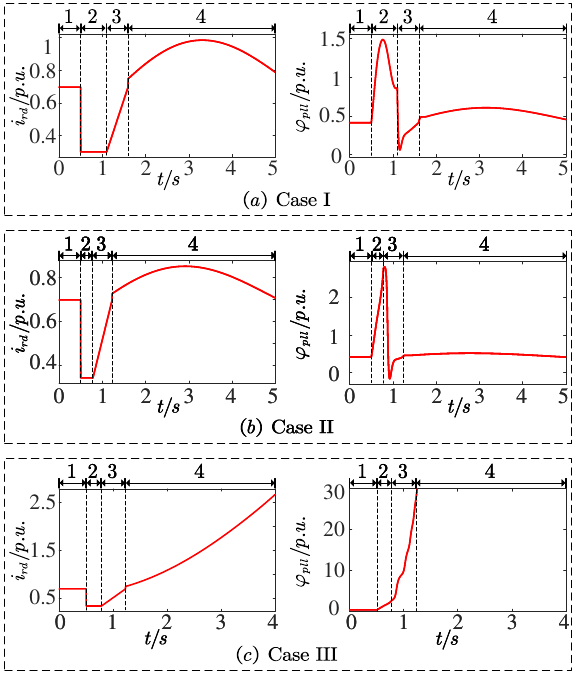}
	\caption{(a)-(c) Time domain simulation results of $i\rm_{\mathit{rd}}$ and $\varphi\rm_{\mathit{pll}}$ for Cases I, II, and III, respectively.}
	\label{Case I}
\end{figure}

\subsection{Simulation verification}			
 					Different cases have been widely studied. Without losing generality, three typical cases with the time domain simulation results of $i\rm_{\mathit{rd}}$ and $\varphi\rm_{\mathit{pll}}$ are shown in \figurename \ref{Case I}.
 					The fault is set as $U_g$ dips to 0.2 p.u. at $t_f$ = 0.5 s, $i\rm_{\mathit{rq}2}=-0.93$ p.u., with all other parameters in Appendix. For the tests, three different values of $i\rm_{\mathit{rd}2}$ and $t_c$ are chosen: 

 					Case I: $i_{\rm\mathit{rd}2}=$0.3 p.u., $t_c$ = 1.1 s, and the  fault duration time is $t_c$ – $t_f$ = 0.6 s.
 													
 					Case II: $i_{\rm\mathit{rd}2}=$0.4 p.u., $t_c$ = 0.782 s, and  $t_c$ – $t_f$ = 0.282 s. 
 												 									
 					Case III: $i_{\rm\mathit{rd}2}=$0.4 p.u., $t_c$ = 0.783 s, and  $t_c$ – $t_f$ = 0.283 s.

 					Clearly in \figurename \ref{Case I}(a), the system is stable in stage 2 and  finally keeps stable in stage 4. In a contrast, from the plot of $\varphi\rm_{\mathit{pll}}$ in \figurename \ref{Case I}(b), one can see that although the system is transiently unstable in stage 2, it becomes stable eventually in stage 4.  
  					Comparing \ref{Case I}(c) with (b), even with a slightly larger $t_c$, the system becomes unstable finally and its fate changes completely. These indicate that even the system is unstable in stage 2, if the fault can be removed in time, the stability can still recover. Clearly these findings imply that the previous EAC-based TSS assessment for permanent faults loses some important information, and the existence of an equilibrium point in stage 2 and the relevant TSS for the device stability in stage 2 are not necessary. 
 			\begin{figure}[!t]
 			\centering
 			\includegraphics[width=0.8 \linewidth]{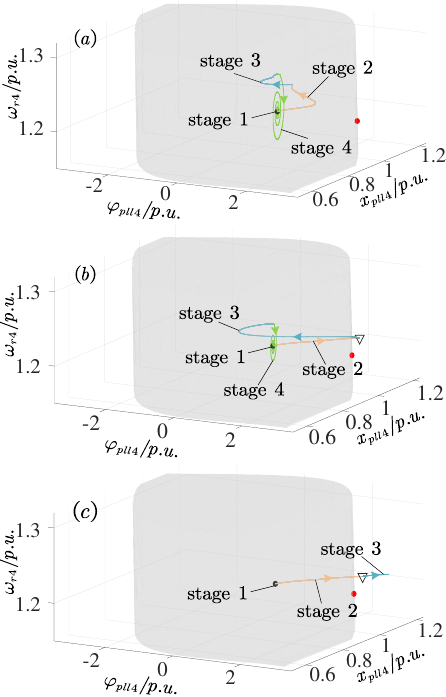}
 				\caption{(a)-(c) Comparisons of BOA in the $\omega\rm_{\mathit{r}}$-$\varphi\rm_{\mathit{pll}}$-$x\rm_{\mathit{pll}}$ three-dimensional space in stage 4 and its fault trajectory for Cases I, II, and III, respectively. A black hollow triangle for the initial state of  stage 3 is superimposed in (b) and (c). 
			}
			\label{3D_BOA}
			\end{figure}
			\begin{figure}[!t]
						\centering
						\includegraphics[width=0.75 \linewidth]{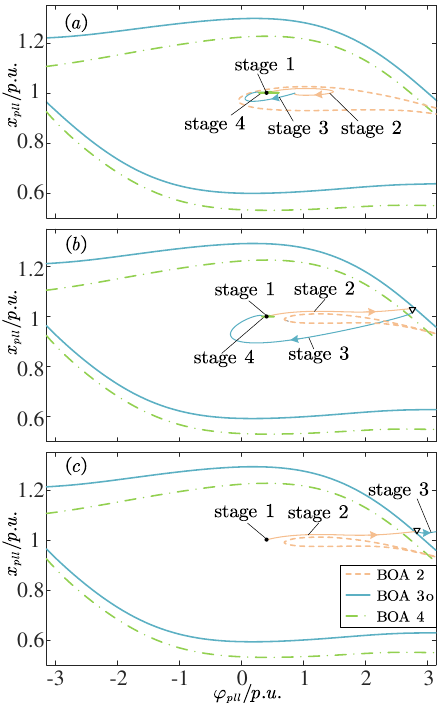} 
						\caption{The same as \figurename \ref{3D_BOA}, but for comparisons on the $\varphi\rm_{\mathit{pll}}$-$x\rm_{\mathit{pll}}$ two-dimensional plane, instead. Similarly, a black hollow triangle for the initial state of stage 3 is superimposed in (b) and (c). This clearly demonstrates that the relation between the initial state of stage 3 and the BOA 3o is dominant in the TSS.
						}
						\label{2D_BOA}	
			\end{figure}

		To show these three cases better, their BOA's in stage 4 and fault trajectories are shown in the $\omega\rm_{\mathit{r}}$-$\varphi\rm_{\mathit{pll}}$-$x\rm_{\mathit{pll}}$ three-dimensional space and the $\varphi\rm_{\mathit{pll}}$-$x\rm_{\mathit{pll}}$ two-dimensional plane in Figs. \ref{3D_BOA} and \ref{2D_BOA}, respectively. As it is difficult to display the BOA in the full five-dimensional space, Fig. \ref{3D_BOA} is a projection, by the other two variables ${i_{\rm{\mathit{rd}}}}$ and ${i_{\rm{\mathit{rq}}}}$ fixed as $i_{\rm{\mathit{rd}1,\mathit{u}}}$ and $i_{\rm{\mathit{rq}1,\mathit{u}}}$ in (\ref{eq_UEP1}), respectively.	
		To emphasize the key difference in Cases II and III induced by the slight parameter change, a black hollow triangle for the initial state of stage 3 is superimposed correspondingly. In Figs. \ref{3D_BOA}(b) and (c), both hollow triangles are out of the BOA of stage 4. 
		Fig. \ref{2D_BOA} clearly shows that the relation between the initial state of stage 3 and the BOA 3o truly plays a determinant role, namely, if it is within (out of) the BOA 3o, the system will be stable (unstable). The TSS can be determined immediately by the initial moment of stage 3. Therefore, with this BOA-based method, the TSS can be well predicted.

\section{EAC-based TSS assessment considering whole LVRT processes}					
\begin{figure}[!t]
				\centering
				\includegraphics[width=.8 \linewidth]{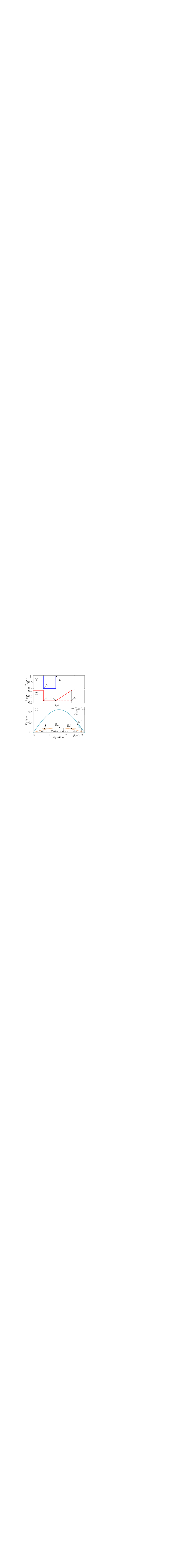}
				\caption{Similar to \figurename \ref{EAC_stage2}, but for the TSS assessment considering whole LVRT processes, instead. For details, see the text.}
				\label{EAC}
				\end{figure}

				After catching the dominant factor in the TSS, it is necessary to extend the EAC-based assessment for permanent faults in Section \ref{sect_stage 2} to the assessment considering the whole LVRT processes. Without losing generality, in \figurename \ref{EAC}(a), a severe fault occurs at $t_f$, $U_g$ dips. The fault is cleared at $t_c$ and $U_g$ is restored. In addition, the variation of 
				$i\rm_{\mathit{rd}2} $ is schematically shown by the solid lines in \figurename \ref{EAC}(b). As the system stability is solely determined by the relation between the initial state of stage 3 and the BOA 3o, for the TSS analysis, the system can be virtually viewed as staying at the stage 3 for ever, i.e., ${i\rm_{\mathit{rd}3}}={i\rm_{\mathit{rd}2}}$ permanently,			
				as shown the added dashed line in \figurename \ref{EAC}(b). In this respect, the only difference with the EAC analysis in \figurename \ref{EAC_stage2} is that the $U_g$'s recovery is incorporated.

				In the EAC analysis in \figurename \ref{EAC}, now ${P_{\rm{\mathit{m}3}}}={P_{\rm{\mathit{m}2}}}$, as the virtual value of $i\rm_{\mathit{rd}3}$ is fixed as ${i\rm_{\mathit{rd}3}} ={i\rm_{\mathit{rd}2}}$. In addition, as $U_{g3}>U_{g2}$, ${P_{\rm{\mathit{e}3}}} > {P_{\rm{\mathit{e}2}}}$. These construct the basic relation between the equivalent mechanical and electromagnetic powers and their change in the EAC analysis. Similarly, the UEP at stage 3, ${\varphi\rm_{\mathit{pll}3,\mathit{u}}}$, is important, 			
				\begin{equation}
				{\varphi\rm_{\mathit{pll}3,\mathit{u}}} =\pi - \arcsin(\dfrac{d{X_g}{i_{\mathit{rd}3}}}{c{U_{\mathit{g}3}}})
				\label{eq_UEPf3}
				\end{equation}

				Therefore, based on the EAC, i.e., the total accelerating area ${S\rm_{\mathit{\Sigma}2}}^{+}$ in stage 2 (starting from 
				${\varphi\rm_{\mathit{pll}1,\mathit{s}}}$) and the maximal decelerating area in stage 3 (ending at ${\varphi\rm_{\mathit{pll}3,\mathit{u}}}$), ${S_3}^{-}$, should be identical, i.e., ${S\rm_{\mathit{\Sigma}2}}^{+} = {S_3}^{-}$, and				
				 \begin{equation}
				 		\label{eq_Ssum}
				 	\left\{ \begin{array}{l}
				 	{S\rm_{\mathit{\Sigma}2}}^{+} = \int_{\varphi\rm_{\mathit{pll}1,\mathit{s}}}^{\varphi_{cr}} ({P_{\rm{\mathit{m}2}}}-{P_{\rm{\mathit{e}2}}}) d{\varphi_{{\rm{\mathit{pll}2}}}}   \\[1mm]					
				    {S_3}^{-} = \int_{\varphi_{cr}}^{{\varphi\rm_{\mathit{pll}3,\mathit{u}}}} ({P_{\rm{\mathit{e}3}}}-{P_{\rm{\mathit{m}3}}}) d{\varphi_{{\rm{\mathit{pll}3}}}}\\[1mm]
				 	\end{array} \right.
				 \end{equation}		 
          Subsequently, the critical clearing angle ${\varphi_{\rm\mathit{cr}}}$ can be calculated, i.e., 
				\begin{align}
						{\varphi_{cr}} = \arccos(\dfrac{  { d{X_g}{i\rm_{\mathit{rd}2}}}     ({\varphi\rm_{\mathit{pll}3,\mathit{u}}}-{\varphi\rm_{\mathit{pll}1,\mathit{s}}})    }{c({U\rm_{\mathit{g}3}}-{U\rm_{\mathit{g}2}})   } \notag \\
					+ \dfrac{{U\rm_{\mathit{g}3}}\cos{{\varphi\rm_{\mathit{pll}3,\mathit{u}}}} - {U\rm_{\mathit{g}2}}\cos{{\varphi\rm_{\mathit{pll}1,\mathit{s}}}} }{ {U\rm_{\mathit{g}3}}-{U\rm_{\mathit{g}2}}}        )												
				\label{eq_phicr}
				\end{align}				
				Next combining the numerical calculation of trajectory, the CCT can be obtained, similar to the TSS in the traditional power systems.

				 As shown in Table I, the CCTs are calculated by the EMT simulation, BOA-based method, and EAC-based method under different values of $U_{\rm{\mathit{g}2}}$ and $i_{\rm{\mathit{rd}2}}$. The relative errors of the two methods are compared with the EMT simulation. 
				 They show that in the BOA-based method the CCTs are always conservative and the relative error is very small. The reason might come from the BOA of stage 3 actually moves, depending on $K\rm_{\mathit{ramp}}$, and is not completely stationary. In addition, the relative error of the EAC-based method is approximately within 13\%, which is still acceptable, and its CCT is always radical. This might come from neglecting the damping term in the EAC. 
				
				Further to demonstrate the influence of $K\rm_{\mathit{ramp}}$, the CCTs with different $K\rm_{\mathit{ramp}}$'s under different methods are studied. $U_{\rm{\mathit{g}2}}=0.2 $ p.u. and $i_{\rm{\mathit{rd}2}}=0.34 $ p.u.. The results are shown in Table II. The CCT calculated by BOA-based method is 0.283s, and that calculated by EAC-based method is 0.270s. As these two methods are based on the initial moment of stage 3, their CCTs are not influenced by $K\rm_{\mathit{ramp}}$. It can be found that as $K\rm_{\mathit{ramp}}$ increases, the CCT of the EMT result only slightly decreases. This is easy to understand; as the BOA moves faster, less time is required to clear faults. In addition, for a larger $K\rm_{\mathit{ramp}}$, the relative error of the BOA-based method increases and oppositely that of the EAC-based method decreases. Nevertheless, the influence of $K\rm_{\mathit{ramp}}$ is tiny.
				\begin{table}[h!]
									\begin{center}
										\label{t1}
										\caption{Comparison of CCT under different values of
										$U_{\rm_{\mathit{g}2}}$ and $i_{\rm_{\mathit{rd}2}}$}
										\scalebox{.96}{
										\setlength{\tabcolsep}{2.56mm}{
										\begin{tabular}{ccccccc}
											\toprule
											\multirow{2}{*}[-1ex]{{\makecell{$U_{\rm_{\mathit{g}2}}$/$i_{\rm_{\mathit{rq}2}}$\\(p.u.)}}} & \multirow{2}{*}[-1ex]{{\makecell{{$i_{\rm_{\mathit{rd}2}}$}\\(p.u.)}}}& \textbf{EMT} & \multicolumn{2}{c}{\textbf{BOA-based}} & \multicolumn{2}{c}{\textbf{EAC-based}}\\
											\cline{3-7}
											& & CCT &CCT &  {\makecell{relative\\error}}  &  CCT&  \makecell{relative\\error} \\  
											\midrule
											\multirow{2}{*}{0.1/-1}    &  0.3 & 0.157s  & 0.158s & + 0.6\%  & 0.143s & - 8.9\% \\  
																   	   &  0.4 & 0.114s  & 0.115s & + 0.9\%  & 0.099s & - 13.2\%\\  
											\cline{1-7}
											\multirow{2}{*}{0.2/-0.93} &  0.34 & 0.282s & 0.283s & + 0.4\%  & 0.270s & - 4.3\% \\  
																	   &  0.5 & 0.124s  & 0.125s & + 0.8\%  & 0.109s & - 12.2\%\\  
											\cline{1-7}
											\multirow{2}{*}{0.3/-0.86} &  0.5 & 0.252s  & 0.253s & + 0.4\%  & 0.239s & - 5.2\%\\  
										    						   &  0.6 & 0.140s  & 0.141s & + 0.7\%  & 0.125s & - 10.7\%\\  
											\bottomrule
										\end{tabular}
										}
										}
									\end{center}
				\end{table}
				
				\begin{table}[h!]
					\begin{center}
						\label{t2}
						\caption{Comparison of CCT under different $K_{\rm_{\mathit{ramp}}}$'s.}
						\scalebox{1}{
						\setlength{\tabcolsep}{2.56mm}{
						\begin{tabular}{cccc}
							\toprule
							\multirow{2}{*}[-2ex]{$K_{\rm_{\mathit{ramp}}}$} & \textbf{EMT} & {\textbf{BOA-based}} & {\textbf{EAC-based}}\\
							\cmidrule{2-4}
							& CCT & \makecell{relative error\\(CCT=0.283s)}	& \makecell{relative error\\(CCT=0.270s)}\\  
							\midrule
							
						0.2-1.2 & 0.282s & + 0.4\%  & - 4.3\% \\  
						1.3-2.9 & 0.281s & + 0.7\%  & - 3.9\% \\  
						3.0-4.5 & 0.280s & + 1.1\%  & - 3.6\% \\  
						4.6-6.1 & 0.279s & + 1.4\%  & - 3.2\% \\  
						6.2-7.8 & 0.278s & + 1.8\%  & - 2.9\% \\  
						7.9-9.6 & 0.277s & + 2.2\%  & - 2.5\% \\  
							\bottomrule
						\end{tabular}
						}
 						}
					\end{center}
				\end{table}

\section{Experimental verification}			
\begin{figure}[h]
	\centering
	\includegraphics[width=0.8 \linewidth]{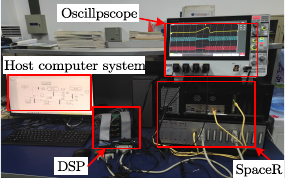}
	\caption{SpaceR real-time simulation platform.}
	\label{HIL}
\end{figure}

In order to verify the above observations, hardware-in-the-loop experiments are conducted based on the SpaceR. The SpaceR real-time simulation platform is shown in \figurename \ref{HIL}. The system model and parameters are the same as in 
\figurename \ref{topology} and the Appendix. Two groups of comparative experiment results are shown here.

\begin{figure}[h]
	\centering
	\includegraphics[width=0.80 \linewidth]{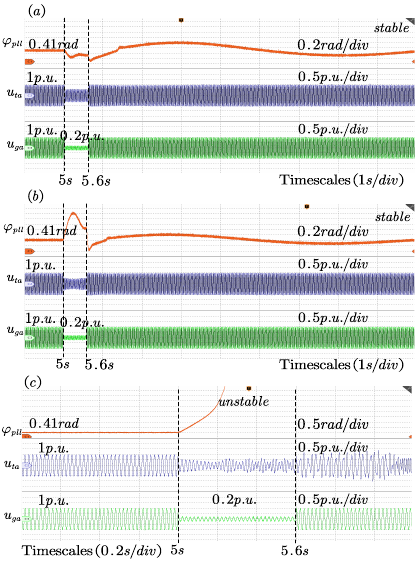}
	\caption{(a)-(c) Experimental waveform diagrams of Case A.}
	\label{Case A}
\end{figure}

Case A: When $t_f$ = 5 s, $U_g$ dips to 0.2 p.u.. At $t_c$ = 5.6 s, $U_g$ recovers to 1 p.u. The experimental results for $i_{\rm\mathit{rd}2}$ = 0.1 p.u., 0.3 p.u., and 0.4 p.u. are shown in Figs. \ref{Case A}(a)-(c), respectively. With increasing of $i_{\rm\mathit{rd}2}$, the risk of transient instability increases. In Figs. \ref{Case A}(a) and (b), when the system is stable in stage 2, it can be stable finally. In \figurename \ref{Case A}(c), when the system is unstable in stage 2, if the fault persists for a longer period of time, the system will eventually lose stability.

\begin{figure}[h]
	\centering
	\includegraphics[width=0.80 \linewidth]{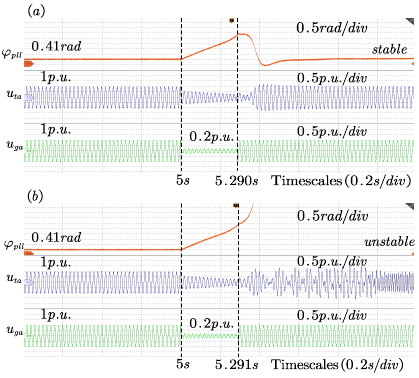}
	\caption{(a) and (b) Experimental waveform diagrams of Case B.}
	\label{Case B}
\end{figure}								
Case B: When $t_f$ = 5 s, $U_g$ dips to 0.2 p.u. and $i_{\rm\mathit{rd}2}$ = 0.34 p.u. At $t_c$, $U_g$ recovers to 1 p.u. The experimental results for two different fault clearing times $t_c$ = 5.290 s and 5.291 s are shown in Figs. 
\ref{Case B}(a) and (b), respectively. It is clear that with a slightly larger $t_c$ in \figurename \ref{Case B}, although $\varphi_{\rm\mathit{pll}}$ is unstable in stage 2, the system can finally become unstable. Under this situation, the CCT is 0.29 s, well in accord with the EMT result: CCT = 0.282 s in Table I. These experimental results demonstrate that even if the system experiences a transient instability in stage 2, as long as the fault is cleared in time, the system can ultimately be stable.

\section{Conclusion}
				In conclusion, the TSS model, analysis, and assessment of the DFIG system considering complete LVRT processes have been systematically studied. The valuable conclusions are as follows:
				
				1) According to the general sequential switching scheme of the LVRT, a transient mechanism model considering complete LVRT processes is constructed. The dominant dynamical behaviors, switching conditions, and chief factors of each stage are studied in detail. Clearly the PLL dynamics is dominant, and the synchronization of DFIG with the grid as a whole can be well characterized by the dynamical behavior of the PLL output angle, which plays a similar role with rotor angle of SG.

				2) By the salient property of the driving-response non-autonomous system of stage 3, the initial moment of stage 3 plays a decisive role in the TSS, namely, if the state is within (out of) the BOA 3o, it will remain within (outside) the BOA 3 and the system will finally be stable (unstable) at stage 4. This point relies on the comparatively slow ramp rate of the DFIG.
				
				3) Standing on the dynamical characteristics of stages 2 and 3, two new TSS assessment methods are developed and compared, including the BOA-based and the EAC-based methods. Their accuracy is well verified by wide simulations and hardware-in-the-loop experiments.
			
				4) The present works not only provide a clear physical picture of the DFIG system and two efficient assessment methods for the TSS but also exhibit guidance for enhancing transient synchronization stability.

				\appendix
				\renewcommand\thefigure{\Alph{subsection}\arabic{figure}}
				
				\setcounter{figure}{0}
				\setcounter{equation}{0}
				\section{}

				Parameters of grid: $S_{\rm\mathit{base}}$ = 2 MW, $U_{\rm\mathit{base}}$ = 690 V (line rms value), $U_{\rm\mathit{dcbase}}$ = 1400 V, $f_0$ = 50 Hz, $\omega_0$ = 2${\pi}f_0$, ${U_{\rm\mathit{dc}}}^{*}$ = 1 p.u., ${U_{\rm\mathit{t}}}^{*}$ = 1 p.u., $U_g$ = 1 p.u., $P\rm_{\mathit{in}}$ = 0.8 p.u., ${\omega_{\rm\mathit{r}}}^{*}$ = 1.2 p.u., $C$ = 0.1 p.u., $L_f$ = 0.1 p.u., $L_g$ = 0.5 p.u. 
				Parameters of DFIG: $L_{\rm\mathit{Is}}$ = 0.171 p.u., $L_{\rm\mathit{Ir}}$ = 0.156 p.u., $L_m$ = 3.9 p.u., $H$ = 4 p.u.
				
				Controller parameters: (1) RSC: $k_{\rm\mathit{pw}}$ = 1, $k_{\rm\mathit{iw}}$ = 5. (2) TVC: $k_{\rm\mathit{pV}}$ = 1, $k_{\rm\mathit{iV}}$ = 10. (3) PLL: $k_{\rm\mathit{ppll}}$ = 60, $k_{\rm\mathit{ipll}}$ = 1400. (4) DVC: $k_{\rm\mathit{pudc}}$ = 3.5, $k_{\rm\mathit{iudc}}$ = 140. (5) ACC: $k_{\rm\mathit{pucd}}$ = 1.3, $k_{\rm\mathit{iucd}}$ = 370, $k_{\rm\mathit{pucq}}$ = 1.3, $k_{\rm\mathit{iucq}}$ = 370, $k_{\rm\mathit{purd}}$ =1.3, $k_{\rm\mathit{iurd}}$ = 370, $k_{\rm\mathit{purq}}$ = 1.3, $k_{\rm\mathit{iurq}}$ = 370. (6) LVRT: $K_e$ = 1.5. (7) Ramp: $K_{\rm\mathit{ramp}}$ = 0.8.


				
				%
				\ifCLASSOPTIONcaptionsoff
				\newpage
				\fi

				
				
				%

				\bibliographystyle{IEEEtran}
				\bibliography{ref}

			\end{document}